# Understanding the Great Recession Using Machine Learning Algorithms

Rickard Nyman[1] and Paul Ormerod[2]

January 2020

The paper formed part of a keynote presentation at the Bank of England/Federal Reserve conference on "Modelling with Big Data and Machine Learning" held at the Bank of England 26-27 November 2018. We are grateful to participants in the conference for comments.

[1] University College London (UCL) and Algorithmic Economics Ltd.; rickard.nyman.11@ucl.ac.uk Corresponding author

[2] University College London (UCL) and Algorithmic Economics Ltd; pormerod@ucl.ac.uk.

*Abstract*

*Nyman and Ormerod (2017) show that the machine learning technique of random forests has the potential to give early warning of recessions.*

*Applying the approach to a small set of financial variables and replicating as far as possible a genuine ex ante forecasting situation, over the period since 1990 the accuracy of the four-step ahead predictions is distinctly superior to those actually made by the professional forecasters.*

*Here we extend the analysis by examining the contributions made to the Great Recession of the late 2000s by each of the explanatory variables. We disaggregate private sector debt into its household and non-financial corporate components.*

*We find that both household and non-financial corporate debt were key determinants of the Great Recession. We find a considerable degree of non-linearity in the explanatory models.*

*In contrast, the public sector debt to GDP ratio appears to have made very little contribution. It did rise sharply during the Great Recession, but this was as a consequence of the sharp fall in economic activity rather than it being a cause.*

*We obtain similar results for both the United States and the United Kingdom.*

# 1. Introduction

As Nyman and Ormerod (2017) note, over the past fifty years or so, a track record of macroeconomic forecasts and their accuracy has been built up on forecasting accuracy, especially with respect to the one year ahead predictions for growth.

A particular problem is the very poor record of predicting recessions. The failure to forecast the financial crisis recession of the late 2000s is well known. But this is a constant theme in the forecasting record over the past 50 years.

For example, four quarters ahead, the mean prediction of the large number of forecasters reported in the Survey of Professional Forecasters has never been for negative growth. In contrast, since 1968 there are 23 examples in the actual data of US real GDP growth being less than zero in a quarter.

Nyman and Ormerod (op.cit.) show that applying a random forest algorithm to a small set of monetary variables generates four-quarter ahead predictions of real US GDP growth which are much better over the 1990-2017 period than those reported by the Survey of Professional Forecasters. Similar results are obtained for the UK.

The machine learning algorithm captures the shallow recession of the early 1990s, the marked slow down (although not technically a recession) in the US economy in the early 2000s after the dotcom crash, and the recession of 2008/09. It does not predict a recession when one did not occur.

In this paper, we extend the previous results on forecasting the Great Recession by examining the contributions made to the Great Recession of the late 2000s by each of the small set of financial variables used in the prediction models. We disaggregate private sector debt into its household and non-financial corporate components.

Section 2 describes the methodology, and section 3 the results.

## 2. Methodology

**2.1 Data selection**

As described in Nyman and Ormerod (op.cit.), we try to replicate as closely as possible an actual forecasting situation.

The dependent variable in our analysis is the third estimate of real GDP growth in the relevant quarter, rather than the most recent estimate which is now available. The third estimate is in general the one on which policy makers would rely when trying to judge the current state of the economy at the time.

Friedman and Schwartz (1963) argued that the Federal Reserve's monetary policies were largely to blame for the severity of the Great Depression. We therefore selected, on this a priori theoretical basis, a small set of monetary variables as potential explanatory factors. We selected them without any preliminary analysis as to whether they would be of any practical use in predicting recessions.

We use the 3 month Treasury Bill rate, the yield on 10 year US government bonds, and the quarterly percentage change in the index of total share prices for all shares in the FRED data base. All these variables are available at the time at which a prediction is made, and their values are not subsequently revised.

We added two private sector debt variables. The ratio of household debt to GDP and the ratio of non-financial corporate debt to GDP. In our previous paper, we used the aggregate of these two, private sector debt to GDP.

There is a small but growing literature within economics which regards debt of individuals (the "household" sector in the jargon) as being of particular importance in the understanding of recessions (for example, Mian and Sufi (2018) and Gertler and Gilchrist (2018)). We therefore disaggregated private sector debt into its component parts.

The debt data is available on the Bank of International Settlements website. This data usually appear with a lag of several months, so the most recent quarter, and possibly the quarter before this, in any forecasting situation would have to have been estimated.

We report results using both the actual values of the two debt variables and the values predicted using an autoregressive process. To anticipate, the results are very similar.

Finally, we added the ratio of public sector debt to GDP to the data set.

**2.2 Selection of algorithm**

Fernandez-Delgado et al. (2014), in a paper whose citations are rising rapidly, compare 179 classification algorithms from 17 "families" such as Bayesian, neural networks, logistic and multinomial regression. They examine their performance on 121 data sets in the University of California at Irvine machine learning repository. This repository is in standard use in machine learning research. The authors find that the random forest family of algorithms achieves the best results. In an economic context, Alessi and Detken (2011) and Alessi et al. (2015) report good results with random forest algorithms in the context of early warning of banking crises.

Random forests (Breiman 2001, 2002) are machine-learning models known for their ability to cope with noisy, non-linear, high-dimensional prediction problems. Many proofs of their properties which extend the original work of Breiman are available in, for example, Biau et al. (2008) and Biau (2012).

They construct a large number of decision trees in training by sampling with replacement from the observations. Each tree in the collection is formed by first selecting at random, at each node, a small group of input coordinates to split on and, secondly, by calculating the best split based on these features in the training set. Each tree gives a prediction, and the predictions are averaged. From the point of view of the bias-variance trade-off, the ensemble of a large number of trees trained on independent bootstrap samples, each with relatively large variance but low bias, achieve much reduced variance without the introduction of additional bias.

More formally, there are a few different variants of random forest classifiers, each built with varying levels of 'randomness'. All variants construct decision trees from samples of the data. A decision tree can be considered a sequence of logical rules applied to a selection of features that ultimately, in the case of classification, assigns an observation to one of the possible classes.

A Random Forest algorithm in its most common form constructs a number of decision trees through a form of bootstrap aggregating (*bagging)*. Several trees are fitted on random samples of the training set $X = x_1, \dots, x_N$ with corresponding class labels $Y = y_1, \dots, y_N$. To minimize the correlation between trees, the data is often sampled with replacement in order for each tree to observe a slightly different set of observations.

Formally, for $b = 1, \dots, B$ (where *B* is the number of trees to be built):
  1. Sample, with replacement, *N* training samples from *X, Y.* Call the samples $X_b, Y_b$

2. Train a classification tree $f_b$ on $X_b, Y_b$

Furthermore, when training each tree $f_b$ a random subset of the features in $X_b$ is considered for each split to ensure that all trees do not use the most predictive features of the training data during construction, to further decrease the correlation between the trees.

When classifying unseen instances each tree makes a classification, and the class mode is selected as the final classification. In the case of regression, the average can be used,

$$\mathcal{F}(x) = \frac{1}{B} \sum_{b=1}^{B} f_b(x)$$

An estimate of the uncertainty of each prediction can be made as the standard deviation of all predictions,

$$\sigma = \sqrt{\frac{\sum_{b=1}^{B} f_b(x) - \mathcal{F}(x)}{B - 1}}$$

Again, we did not experiment with different rolling windows for the training period, but simply moved it forward one quarter at a time.

Finally, we report the average of 100 separate predictions made with the random forest algorithm.

## 3. Analysis and Results for the United States

### 3.1 The prediction framework

We used the statistical program R, and downloaded the package *randomForest* to carry out the random forest analysis. We also checked the results by using the Python *SciKit-Learn* library. The two sets of results are virtually identical, and we report here the ones obtained in R.

We emphasise that we used the default values for the various options available for inputs in the random forest algorithm. In other words, we did not attempt in any way to optimise the accuracy of the predictions by trying different combinations of input parameters.

Further, we only used the explanatory variables from the final quarter of the estimation period and from the quarter previous to this in predicting GDP growth four quarters ahead. So, for example, in making a prediction of real GDP growth in 1990Q2, say, we used values of the explanatory variables in 1989Q1 and 1989Q2.

We initially trained a model over the period 1970Q2 to 1989Q2. For the four-step ahead forecast, we predicted 1990Q2. We then trained the model 1970Q2 to 1989Q3 and predicted 1990Q3, and so on until we predicted 2010Q4, by which time the US economy had clearly emerged from the Great Recession of 2008/09.

**3.2 A benchmark comparator**

In Nyman and Ormerod (op.cit.), we generated a comparison benchmark by regressing actual growth of real GDP on the mean of the forecasts reported in the SPF. The use of the median instead of the mean gives slightly worse results. This is because it gives less weight than the mean at times of recession to any outliers in the forecasting community which may have happened to predict negative growth.

We used both the third estimate of GDP growth and its most recent estimate. The results were very similar.

Essentially, for one period ahead predictions, the SPF explains around 25 per cent of the variability in the actual data. The precise amount varies depending on the estimation period chosen, but it is never very much above this.

But over a four-quarter ahead horizon, the SPF track record explains none (or very little over certain periods) of the variability of the data.

### 3.3 Results using random forest algorithm

Initially, we used data on Treasury Bills, 10 year bonds, percentage change in share prices, public sector debt to GDP ratio, household debt to GDP ratio and non-financial corporate debt to GDP ratio.

In predicting 1990Q2, we used data for these variables in 1989Q1 and 1989Q2.

We did not experiment with longer lags, or with omitting one of the lags on a variable-by-variable basis. In other words, we retained this simple lag structure.

The results in the table are obtained when data on the 10 year bond and the public sector debt ratio variable are omitted from the set of explanatory variables. The inclusion of either or both of these variables gave results which were almost identical to those in the table.

In addition, we report results using both the actual data on household and corporate debt, and two-quarter ahead recursive predictions generated by a simple AR(2) model (i.e. the predicted value one quarter ahead is used in the prediction for the second quarter ahead). As discussed above, data on these variables appear with a time lag.

Before setting out the table of results, the final point to make is that these are the average of 100 separate predictions made using the random forest algorithm.

We report the regression of the actual third estimate of GDP growth on the random forest predictions over the four quarter horizon in Table 2 below.

**Table 2** Regressions of GDP quarter on quarter growth, annualised rate, per cent, third estimate on the average of 100 random forest predictions made four quarters previously, 1990Q2- 2010Q4

Dependent variable: GDP quarter on quarter growth, annualised rate, per cent, third estimate

|  | Using actual values of the debt variables | Using predicted values of the debt variables |
|---|---|---|
| Prediction | 0.824 | 0.794 |
|  | (0.157) | (0.152) |
| Constant | 0.304 | 0.442 |
|  | (0.500) | (0.481) |
| Adjusted R2 | 0.246 | 0.243 |
| Residual Std. Error | 2.146 | 2.151 |

Compared to the explanatory power of zero of the SPF forecasts, the random forest model obtains a degree of power similar to that of the one quarter ahead SPF forecasts.

Using the same methodology with the same set of variables, but using Ordinary Least Squares, we obtain an R bar squared of only 0.025 using actual values of the debt variable and 0.032 using the values predicted by an AR(2) model as described above. This implies that, certainly at key turning points, there is considerable non-linearity in the explanatory relationship.

The table below shows the actual third estimate GDP growth and the predictions made by the random forest approach four quarters previously, over the 1990Q2-2010Q4 period.

Intriguingly, although the random forest predictions would not have got the exact timing of the recession in the winter of 2008/09 correct, a serious recession would have been predicted for early 2009 eighteen months previously.

| Quarter | Actual annualised third estimate quarter on quarter real GDP growth | Random forest predictions made four quarters previously, using estimated data for debt variables |
| --- | --- | --- |
| 2008Q1 | 0.96 | 2.33 |
| 2008Q2 | 2.83 | 2.02 |
| 2008Q3 | -0.51 | 2.49 |
| 2008Q4 | -6.34 | 1.17 |
| 2009Q1 | -5.49 | -2.54 |
| 2009Q2 | -0.74 | -3.10 |
| 2009Q3 | 2.24 | -2.46 |
| 2009Q4 | 5.55 | 1.34 |

The results are really quite striking. For example, in 2008Q1, a prediction of a negative growth rate of -2.54 per cent in 2009Q1 could have been made. The overall depth of the recession predicted for 2009Q1-2009Q3 was obviously not as big as the actual recession itself, but it is very similar to the recession of the early 1980s, which until 2008/09 was distinctly the largest recession experienced since the 1930s.

There are two other periods of recession in the period on which we focus: the winter of 1990/91 and 2001 (technically, there was only one period of negative growth in the third estimate data for 2001, but growth was close to zero in other quarters). In both cases, the random forest approach predicts four quarters ahead a marked slowdown in growth, but again slightly later than occurred. The approach does not predict a recession, but several periods of growth under 1 per cent (at an annualised rate). In general, periods with growth at this low level are associated, for example, with rising unemployment.

The random forest does *not* predict a recession in periods when one did not in fact take place.

## 3.4 Uncertainty around the 100 random forest predictions

Within each random forest, 500 separate decision trees are constructed, 500 being the default value for the number of trees in the R package which we used. As noted in section 2.2 above, for any given forecast for an individual random forest, a measure of uncertainty can be constructed. This is based on the spread of the individual tree predictions, in this case 500.

For the forecasts whose performance is summarised in Table 2 above, we can average these spreads across the 100 separate random forests for each data point in the forecast period.

The summary statistics for the standard deviation are as follows for the period 1990Q2 to 2007Q4:

Min: 0.99; 1$^{st}$ quartile: 1.44; mean: 1.79; 3$^{rd}$ quartile: 2.05; max: 3.31

The correlation between the predicted growth in GDP and the standard deviation around the random forests is not significantly different from zero over this period.

When we drill down to the individual quarters over the 2008Q1 to 2009Q4 period, we can readily see that the spread around the tree predictions widens sharply at the turning point.

| Quarter | Predicted GDP Growth | Average of the standard deviations within each of 100 random forests |
|---|---|---|
| 2008Q1 | 2.33 | 2.08 |
| 2008Q2 | 2.02 | 1.91 |
| 2008Q3 | 2.49 | 1.54 |
| 2008Q4 | 1.17 | 1.68 |
| 2009Q1 | -2.54 | 4.35 |
| 2009Q2 | -3.10 | 4.54 |
| 2009Q3 | -2.46 | 3.69 |
| 2009Q4 | 1.34 | 5.08 |

This illustrates an interesting feature of random forest predictions. Any individual tree within any given forest might at a given time generate a very poor forecast. It is the averaging of the trees which is a key feature of the random forest approach.

At turning points, the variability across the trees within an individual forest tends to rise sharply. But, again, the averaging produces a reasonable prediction.

## 3.5 Relative importance of the variables

Mullainathan and Spiess (2017) point out that machine learning algorithms typically revolve around the question of prediction, whilst the main focus of econometrics is parameter estimation.

The ability to identify parameters familiar to econometricians from results based on machine learning is still an unsolved problem. Athey and Luca (2019) discuss the question and note that progress is being made (for example, Wager and Athey (forthcoming)).

One approach which has been applied quite widely is known as LIME, the acronym used for "Locally Interpretable Model Agnostic Explanations" (Ribeiro et al.2016). At the risk of over-simplification, this essentially uses linear approximations in any given neighbourhood of a prediction.

The problem we found with applying this technique to the period of the Great Recession, for example, was that by far the strongest correlation was between the Treasury Bill rate and the model predictions of GDP. This itself might not be a problem. The issue was that the linear approximation gave a positive correlation between the two variables. The rate was 3.4 per cent in 2007Q4 and was cut to only 0.3 per cent in 2008Q4. By 2009Q4 it was effectively zero.

We therefore used a rough and ready approach by omitting each of the potential explanatory factors at a time, and also certain combinations of them. We examined the effect on the forecasting performance.

The results are set out in Table 3 below, showing the adjusted R-squared values. These values are the average of each of the R-squared values across 100 separate random forest predictions.

**Table 3  Adjusted R squared values of regressions of actual GDP quarter on quarter growth, annualised rate, per cent, third estimate on the average of 100 random forest predictions made four quarters previously, using estimates of debt variables, 1990Q2-2010Q4**

| | |
|---|---|
| Including Treasury Bill, change in share prices, Household and Corporate debt | 0.243 |
| Omitting change in share prices | 0.228 |
| Omitting Treasury Bill rate | 0.223 |
| Omitting household debt to GDP | 0.203 |
| Omitting corporate debt to GDP | 0.246 |
| Omitting both debt variables | 0.116 |

These results are not necessarily conclusive, but suggest a decisive role for private sector debt as a cause of the Great Recession.

## 4  Results for the United Kingdom

We describe these results considerably more briefly, adopting the same approach as above. There is no historical record of actual forecasts made which is similar to that of the SPF in the US, so we are unable to make this direct comparison.  But, in general, the macroeconomic forecasting record is similar in the two countries.

As explanatory variables, we use the 3-month Treasury Bill rate, the yield on 10 year UK government bonds, the quarterly percentage change in the index of total share prices for all shares in the FRED data base and the ratios of public, household and corporate debt to GDP.

Again, we train the model initially over the 1970Q2 – 1989Q2 period, and predict 1990Q2, moving the analysis and prediction of the Office for National Statistics third estimate of real GDP[3] one quarter at time through to 2010Q4.

Initial screening suggested the omission of the two interest rate variables, so the results reported use the three debt variables and the change in share prices.

Looking four quarters ahead, the OLS generated predictions give an R bar squared for the regression of actual GDP growth on its predicted value of 0.093.  In contrast the R bar squared using the predictions from the random forest model is 0.293.   Again, considerable non-linearities appear to exist in the data.

Leaving out each of the explanatory variables in turn leads to results which are similar to those for the United States.

Omitting the change in share prices from the explanatory data set leads the adjusted R squared to fall from 0.293 to 0.265.

Omitting public debt to GDP from the explanatory data set leads the adjusted R squared to fall from 0.293 to 0.274.

Omitting the change in household debt from the explanatory data set leads the adjusted R squared to fall from 0.293 to 0.258.

---

[3] The GDP data prior to1997 is in fact the latest estimate not the third, the relevant database only giving the third estimate back to 1997Q1.

Omitting the change in corporate debt from the explanatory data set leads the adjusted R squared to fall from 0.293 to 0.292.

But, again, leaving out both private sector debt variables leads the adjusted R squared to fall to 0.092

## 5 Concluding remarks

We have tried, as far as it is possible, to replicate an actual forecasting situation starting for the United States in 1990Q2 and moving forward a quarter at a time through to 2010Q4. We use a small number of lags on a small number of financial variables in order to make predictions.

In terms of one step ahead predictions of real GDP growth, we have not been able to improve upon the mean forecasts made by the Society of Professional Forecasters.

However, even just four quarters ahead, the SPF track record is very poor. A regression of actual GDP growth on the mean prediction made four quarters previously has zero explanatory power, and the SPF predictions never indicated a single quarter of negative growth. The random forest approach improves very considerably on this.

To emphasise, we have not attempted in any way to optimise these results in an *ex post* manner. We use only the default values of the input parameters into the machine learning algorithm, and use only a small number of explanatory variables.

We obtain qualitatively similar results for the UK.

The evidence suggests quite clearly that public sector debt played no causal role in generating the Great Recession. Omitting this variable from the explanatory set in both the US and the UK makes no difference to the predictive power of the model.

In contrast, the ratios of household debt to GDP and non-financial corporate sector debt to GDP does appear to have played a significant role.

As Ormerod and Mounfield (2000) show, using modern signal processing techniques, the time series GDP growth data is dominated by noise rather than by signal. So there is almost certainly a quite restrictive upper bound on the degree of accuracy of prediction which can be achieved.

However, machine learning techniques do seem to have considerable promise in extending useful forecasting horizons and providing better information to policy makers over such horizons.

# References

Alessi, L. and Detken, C., 2011. Quasi real time early warning indicators for costly asset price boom/bust cycles: A role for global liquidity. *European Journal of Political Economy*, *27*(3), pp.520-533

Alessi, L., Antunes, A., Babecký, J., Baltussen, S., Behn, M., Bonfim, D., Bush, O., Detken, C., Frost, J., Guimaraes, R. and Havranek, T., 2015. Comparing different early warning systems: Results from a horse race competition among members of the macro-prudential research network.

Athey, S. and Luca, M., 2019. Economists (and Economics) in Tech Companies. *Journal of Economic Perspectives*, *33*(1), pp.209-30

Biau, G., 2012. Analysis of a random forests model. *Journal of Machine Learning Research*, *13*(Apr), pp.1063-1095.

Biau, G., Devroye, L. and Lugosi, G., 2008. Consistency of random forests and other averaging classifiers. *Journal of Machine Learning Research*, *9*(Sep), pp.2015-2033.

Breiman, L., 2001. Random forests. *Machine learning*, *45*(1), pp.5-32

Breiman, L., 2002. Manual on setting up, using, and understanding random forests v3. 1. *Statistics Department University of California Berkeley, CA, USA*

Fernández-Delgado, M., Cernadas, E., Barro, S. and Amorim, D., 2014. Do we need hundreds of classifiers to solve real world classification problems. *J. Mach. Learn. Res*, *15*(1), pp.3133-3181.

Fildes, R. and Stekler, H., 2002. The state of macroeconomic forecasting. *Journal of Macroeconomics*, *24*(4), pp.435-468

Friedman, M. and Schwartz, A., 1963. *A monetary history of the United States*

Mullainathan, S. and Spiess, J., 2017. Machine learning: an applied econometric approach. *Journal of Economic Perspectives*, *31*(2), pp.87-106.

Ormerod, P. and Mounfield, C., 2000. Random matrix theory and the failure of macro-economic forecasts. *Physica A: Statistical Mechanics and its Applications*, *280*(3), pp.497-504.

Ribeiro, M.T., Singh, S. and Guestrin, C., 2016, August. Why should I trust you?: Explaining the predictions of any classifier. In *Proceedings of the 22nd ACM SIGKDD international conference on knowledge discovery and data mining* (pp. 1135-1144). ACM.

Stockton, D., 2012. *Review of the Monetary Policy Committee's Forecasting Capability,* presented to the Court of the Bank of England
http://www.bankofengland.co.uk/publications/Documents/news/2012/cr3stockton.pdf

Varian, H.R., 2014. Big data: New tricks for econometrics. *The Journal of Economic Perspectives*, *28*(2), pp.3-27

Wager, S., and Athey, S., forthcoming. Estimation and Inference of Heterogeneous Treatment Effects using Random Forests." *Journal of the American Statistical Association* 109(508).

Zarnowitz, V. and Braun, P., 1993. Twenty-two years of the NBER-ASA quarterly economic outlook surveys: aspects and comparisons of forecasting performance. In *Business cycles, Indicators and Forecasting* (pp. 11-94). University of Chicago Press.